\documentclass[aps,prd,10pt,twocolumn,groupedaddress,showpacs,floatfix]{revtex4-1}
\usepackage{textcomp,gensymb}
\usepackage{hyperref}
\usepackage{graphicx}
\usepackage{amsfonts,amsmath,amssymb,bm,bbm}
\usepackage{color,soul}
\usepackage{slashed}
\usepackage[toc,page]{appendix}
\usepackage{flushend}
\usepackage{physics}
\usepackage{graphicx}
\usepackage{dcolumn}
\usepackage{float}

\hypersetup{
    pdfnewwindow=true,      
    colorlinks=true,       
    linkcolor=blue,          
    citecolor=blue,        
    filecolor=blue,      
    urlcolor=blue        
}

\newcommand{\eq}[1]{eq.~(\ref{eq:#1})}

\newcommand{\sect}[1]{section~\ref{sec:#1}}

\begin{document}

\title{Dark matter boosted by terrestrial collisions}

\author{Zamiul Alam}

\author{Christopher V. Cappiello}

\author{Francesc Ferrer}

\affiliation{{Department of Physics and McDonnell Center for the Space Sciences,
Washington University, St. Louis, Missouri 63130, USA}}

\begin{abstract}
	Inelastic dark matter (IDM) models feature an energy threshold for scattering with Standard Model particles, 
	which enables their consistency with the increasingly stringent limits placed by direct detection experiments. 
	In a typical construction, elastic scattering is absent at tree level, and a lighter dark matter state must 
	first upscatter into a heavier state in order to interact with the nuclei in the detector.
	We model the excitation of IDM in the Earth followed by its downscattering inside a detector, and we show that considering this process markedly enhances the sensitivity of 
	existing detectors. In particular, current limits based on XENON100 and XENON1T data can be extended to significantly larger mass splittings.
\end{abstract}

\maketitle

\section{Introduction}

Observations on astrophysical and cosmological scales reveal the gravitational effects of large amounts of dark matter (DM) 
contributing about one fourth of the critical density in the universe 
or $\sim 85$\% of the total matter density~\cite{ParticleDataGroup:2024cfk}. 
Any fundamental particle constituting the DM must be part of an extension beyond the Standard Model (SM) of 
particle physics and it is generically expected to have other interactions, besides gravity, 
that could be probed in the laboratory. 
In this respect, an attractive class of DM candidates is made of particles with masses in the 
range few GeV $ < m_\chi <$ few TeV and weak-scale interactions. These particles, that include thermal relics like
WIMPs, may be probed in direct detection experiments that monitor their low-energy scattering off target nuclei in the laboratory. 

Decades of searches in ultra-sensitive low-background facilities have set strong limits, but have yet to produce a conclusive 
detection~\cite{Bertone:2016nfn,Schumann:2019eaa,Lin:2019uvt,Cooley:2022ufh}. There are many possible reasons for this lack of detection. 
Dark matter may be too weakly interacting to discover with current searches~\cite{Battaglieri:2017aum, Leane:2018kjk, Akerib:2022ort}. 
It may be so light that it does not trigger most conventional detectors~\cite{Kahn:2021ttr,Adams:2022pbo, Mitridate:2022tnv, Zurek:2024qfm}, 
or alternatively, it could be so heavy that it rarely if ever passes through such 
detectors~\cite{Carr:2020xqk,Green:2020jor,Blanco:2021yiy,Carney:2022gse}. 
Another possibility is that dark matter interactions with SM particles
could be predominantly \emph{inelastic}, having an energy threshold that suppresses or entirely forbids scattering with laboratory 
targets at typical Milky Way 
velocities~\cite{Hall:1997ah, Tucker-Smith:2001myb, Tucker-Smith:2004mxa, Arina:2007tm, Cui:2009xq, Feldstein:2010su, Fox:2010bu, An:2011uq, Pospelov:2013nea, Dienes:2014via, Barello:2014uda, Eby:2019mgs, Baryakhtar:2020rwy, Bell:2020bes, Harigaya:2020ckz, Bell:2021xff, Emken:2021vmf, Alves:2023kek}.

In a typical inelastic dark matter scenario (iDM), most of the dark matter in the universe is in the form of a (relatively) light particle $\chi_1$ with mass $m_1$. At tree level, $\chi_1$ cannot scatter elastically with Standard Model particles, but instead must be excited into a higher mass state $\chi_2$.
Although elastic scattering may be possible at loop level, in this work we assume that this process is heavily suppressed and it can be neglected. As a result, in order for dark matter to scatter with a nucleus or electron, the total kinetic energy in the CM frame must be at least equal to the mass splitting $\delta \equiv m_2 - m_1$. If the mass splitting is too large, or the kinetic energy too small, tree-level scattering is thus forbidden, and the dark matter can evade direct detection searches.

The previous discussion implicitly assumes that in order to observe iDM via 
direct detection, the dark matter must upscatter inside the detector, 
scattering off a xenon nucleus for example. 
In this work we study the possibility that iDM upscatters in the Earth outside of the laboratory, and 
subsequently downscatters on xenon inside a direct detection experiment. As we discuss below, 
taking into account this possibility enables us to probe additional previously unexplored regions in
iDM parameter space. 
As a matter of fact, the inelastic conversion of $\chi_1$ into $\chi_2$ in 
the Earth has been previously considered in the literature. 
However, for the subsequent signal on the detector it was assumed that $\chi_2$ either decayed 
electromagnetically~\cite{Feldstein:2010su,Pospelov:2013nea,Eby:2019mgs,Bell:2020bes,Eby:2023wem,Graham:2024syw}, 
or that it would scatter on target electrons via a dark 
photon~\cite{Emken:2021vmf}. We present a more generic scenario, in which both 
the excitation and deexcitation occur through the same mechanism, namely 
inelastic scattering with nuclei. 
Due to the different kinematics of upscattering and downscattering, we are 
able to probe mass splittings that are substantially 
higher than those constrained by past direct detection searches, while using 
the same  
data and remaining relatively model-independent. 
We first present limits based on XENON1T~\cite{XENON:2018voc}, and then we
consider a
XENON100 re-analysis reaching much larger recoil energies~\cite{XENON:2017fdd}.

The remainder of the paper is organized as follows. In \sect{upscattering} we 
review the kinematics of iDM scattering, including the differences between upscattering and downscattering. 
In \sect{upearth} we calculate the expected event rate in a detector if, due 
to a previous interaction in the Earth, iDM arrives at the laboratory in the 
excited state $\chi_2$. Our results are presented in \sect{results}, and
finally, we discuss our findings and present our conclusions 
in \sect{conclusions}.

\section{Kinematics of inelastic dark matter collisions}
\label{sec:upscattering}

We assume that effectively all cosmological dark matter exists in the form of a particle $\chi_1$ with mass $m_1$. 
When this particle scatters off a nucleus, it may be excited to a second state $\chi_2$ with mass $m_2 = m_1 +\delta$, 
where $0 < \delta \ll \{m_1, m_2\}$. 
$\chi_2$ can also downscatter on nuclei, returning to the $\chi_1$ state. 
Since we are taking the mass splitting $\delta$ to be small, we will sometimes use 
$m_{\chi}$ to denote either $m_1$ or $m_2$, in situations where the difference between them is 
irrelevant. Throughout this work, we take $m_{\chi}$ = 1 TeV, motivated by similar work on 
higgsinos and for ease of comparison with related work on more general models of 
inelastic dark matter~\cite{Pospelov:2013nea,Bramante:2016rdh,Eby:2019mgs,Song:2021yar,Eby:2023wem,Graham:2024syw}.
Additionally, we assume that the states $\chi_{1,2}$ couple to 
nuclei 
via the standard spin-independent interaction.

When $\chi_1$ scatters with a nucleus (of mass $m_A$) at rest, energy conservation dictates that
\begin{equation}
    \frac{1}{2}m_1v_i^2 = \frac{1}{2}m_2v_f^2 + \frac{1}{2}m_Av_A^2 + \delta\,,
	\label{eq:econs}
\end{equation}
where $v_i$ and $v_f$ are the initial and final dark matter velocities, and $v_A$ is the final velocity of the nucleus. 
Clearly there is an energy threshold for this reaction,
which can be translated into a threshold value of $v_i$~\cite{Song:2021yar}:
\begin{equation}
    	v_{\textrm{th}} = \sqrt{\frac{2\delta}{\mu_{\chi A}}}\,,
\end{equation}
with $\mu_{\chi A}$ being the reduced mass of the nucleus-iDM system.

In contrast, downscattering has no energy threshold, and it 
can proceed unhindered at an arbitrarily low iDM velocity.
However, rewriting \eq{econs} in terms of momentum transfer 
$q = \sqrt{\pqty{m_2 \mathbf{v}_f- m_1 \mathbf{v}_i}^2}$ and scattering 
angle $\theta = \angle\pqty{\mathbf{v}_i,\mathbf{v}_f}$
shows that, for both upscattering and downscattering, there is a
a minimum and a maximum value of $q$, which is attained at 
$\vqty{\cos\theta}=1$. It then follows that there is a 
minimum velocity required 
to produce a nuclear recoil with a specific energy 
$E_r = q^2/(2 m_A)$~\cite{Emken:2021vmf,Song:2021yar}:
\begin{equation}
	v_{\text{min}}(E_r) = 
	\left|\frac{m_AE_r}{\mu_{\chi A}} \pm \delta \right|\frac{1}{\sqrt{2m_AE_r}}\,,
    \label{eq:vminfunction}
\end{equation}
where the positive sign corresponds to upscattering and the negative sign to downscattering.

Unlike searches for light DM, which lose sensitivity if an experiment cannot 
detect low enough recoil energies, searches for inelastic DM are often 
limited by not probing high enough energies. Indeed, from \eq{vminfunction} 
above we see that if $\delta$ is the larger of the two terms in the bracket, 
then the minimum velocity required to produce a recoil of energy $E_r$ 
decreases as $E_r$ increases. 
Because mass splittings of a few hundred keV typically require velocities near the maximum DM velocity, 
searching for higher energy recoils can increase the event rate or extend the sensitivity to higher values of
$\delta$~(see~\cite{Bramante:2016rdh,Song:2021yar} for a detailed illustration of this point).

However, as we will see in the next section, the sign flip in \eq{vminfunction} allows us to probe larger mass splittings 
for a given recoil energy if the iDM upscattering does not take place in the detector, but the downscatter does.

\section{Inelastic dark matter upscattering in the Earth}
\label{sec:upearth}

We now consider the possibility that iDM arrives at the detector in the
form of an excited state $\chi_2$, due to a previous upscattering of 
$\chi_1$ in the Earth. We take the local dark matter density to be 
0.4 GeV/cm$^3$, and we assume that all iDM in the Galaxy
is initially in the ground state $\chi_1$.

\subsection{Target Nuclei}

When dark matter scatters with a nucleus, the center-of-mass kinetic energy depends on the reduced mass of the dark matter-nucleus system. Because we consider dark matter much heavier than a nucleus ($m_{\chi} = 1$ TeV), the kinetic energy in the center-of-mass frame is larger for heavier nuclei. Therefore, the heavier the target nucleus is, the easier it is to upscatter the dark matter. While the Earth is largely composed of relatively light nuclei such as oxygen and silicon, a small but non-negligible fraction of the crust and mantle is composed of lead. Because we are primarily concerned with large mass splittings, in this work we mostly consider dark matter upscattering on lead.

\subsection{Dark matter velocity}

To estimate the expected signal at the underground detector we need
to specify the initial velocity of $\chi_1$ when it first hit a nucleus
of the Earth.

To first order, one usually assumes that the dark matter particles have
velocities drawn from a Maxwell-Boltzmann distribution~\cite{Drukier:1986tm} with dispersion
$\sigma \sim 220$ km/s and a high-velocity cutoff set by
the escape velocity from the Galaxy $v_{esc} \sim 540$ km/s (see, e.g., Ref.~\cite{XENON:2018voc}). After boosting into the rest frame of the Earth, the peak velocity gets shifted to around 300 km/s, and the maximum velocity is increased to around 800 km/s (see, e.g., Ref.~\cite{Evans:2018bqy, Besla:2019xbx}). However, this model is known to be an approximation. For example, observations from \emph{Gaia} have provided evidence for both a somewhat lower escape velocity~\cite{2024ApJ...972...70R} and a more complicated functional form for the velocity distribution~\cite{Evans:2018bqy, Necib:2018iwb}.

Interestingly, it was argued in Ref.~\cite{Besla:2019xbx} that the 
high-velocity tail of the DM distribution in the Milky Way should be 
dominated by DM particles falling into the Galaxy from the Large Magellanic 
Cloud (LMC). This idea was further developed in Ref.~\cite{Smith-Orlik:2023kyl}, using magneto-hydrodynamic simulations of Milky Way analogs. 
In the present work, we use a velocity distribution for dark matter drawn directly from the numerical results of Ref.~\cite{Smith-Orlik:2023kyl}. Because of the need for high velocity to excite large mass splittings, the enhancement from the LMC will help extend our results to large $\delta$. 

\begin{figure}[t]
    \centering
    \includegraphics[width=\linewidth]{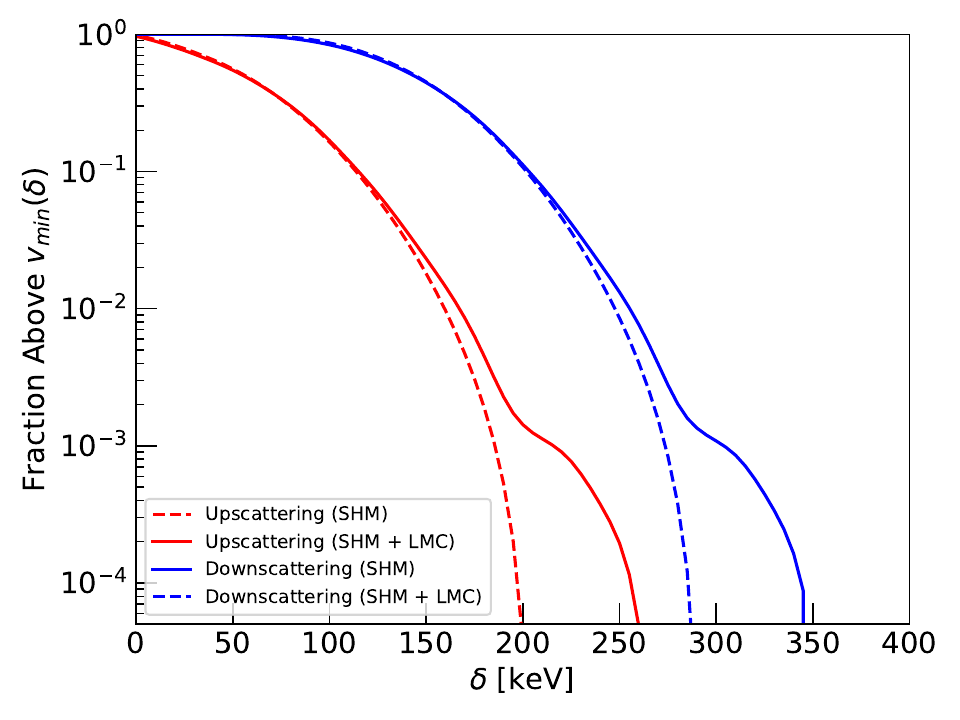}
    \caption{Fraction of DM particles capable of producing a xenon recoil between 4.9 and 40.9 keV. The red curves are for upscattering, while the blue curves are for downscattering. The dashed curves are including only the Standard Halo Model (SHM), while the solid curves include the contribution from the Large Magellanic Cloud (LMC)~\cite{Smith-Orlik:2023kyl}.}
    \label{fig:fractions}
\end{figure}

In Fig.~\ref{fig:fractions}, we show the fraction of DM particles capable of producing a xenon recoil in the energy range between 4.9 keV and 40.9 keV, the analysis window used in Ref.~\cite{XENON:2018voc}. This plot illustrates two things. First, the high-velocity shoulder due to the LMC is clearly visible. Although the fraction of DM particles coming from the LMC is small, they dominate the high-velocity tail of the distribution, and this high velocity maps directly onto the maximum value of $\delta$ for which $\chi_1$ can be excited to the $\chi_2$ state. Second, if the incoming flux is composed entirely of $\chi_2$, then the mass splitting that can be probed is larger by nearly 100 keV than if the flux is entirely composed of $\chi_1$. This motivates our search for DM that has been upscattered into the excited state before being detected.

\subsection{Event Rate}

When a $\chi_1$ particle passes through the Earth, it may collide with a lead nucleus and upscatter to $\chi_2$. The probability for this process to occur in a finite distance $L$ is simply $P = n_\text{Pb} \sigma_{\chi A} L$ (as long as that probability is $\ll$ 1), where $n_\text{Pb}$ is the number density of lead and $\sigma_{\chi A}$ is given by:
\begin{equation}
    \sigma_{\chi A} (v) = \int_{0}^{\infty}
\frac{d \sigma}{d E_R} d E_R \,\Theta \big(v - {v_{\text{min}}} (E_R)\big)\,,
\end{equation}
with
\begin{equation}
    \frac{d\sigma}{dE_{R}} = \frac{m_A \sigma_p}{2 \mu_{\chi p}^2 v_\chi^2} A^2 \left| F_A(q) \right|^2 \bigg|_{q = \sqrt{2 m_A E_R}}\,.
\end{equation}
Here the factor $\Theta \big(v - v_{\text{min}}(E_R)\big)$ enforces the bounds of the energy integral, and $F_A(q)$ is the Helm form 
factor~\cite{Helm:1956zz,Duda:2006uk}. For the crust, we use a lead number density of 1.22$\times$10$^{18}$ cm$^{-3}$~\cite{2003TrGeo...3....1R}, while for the mantle we use 2.8$\times$10$^{16}$ cm$^{-3}$~\cite{2003TrGeo...2....1P}.

One key difference between our approach and a typical inelastic DM search is that we do not care about the recoil energy of the lead. Experimental searches typically look for upscattering in a detector, meaning they must be sensitive to large recoil energy in order to probe large mass splittings. But in our work, all that is needed in the initial collision is for the DM to become excited. 

We assume that the DM arrives at the detector isotropically, and neglect its change in direction when scattering with lead (as the DM is much heavier than a lead nucleus). The resulting distances that DM travels through the Earth range from $\sim$1 km (the depth of the XENON detectors) to $\sim12,700$ km (the diameter of Earth). We sum over 20 different incoming angles distributed isotropically, with each angle representing a different trajectory through the Earth. For simplicity, for DM that arrives from below the horizon, we neglect contributions from the crust and treat the entire trajectory as passing through the mantle. 
We also assume for simplicity that the core has the same lead density as the mantle (the fraction of particles that pass through the core is $<$10\%, so detailed modeling of the core has a negligible effect on our results).

Once the DM has been excited, we compute the rate of it downscattering in the detector. This computation is done in almost the same way as a standard direct detection analysis, with two changes. First, there is a velocity threshold, as in the upscattering process. And second, we must account for the fact that the DM has slowed down slightly in the upscattering process. With both of these effects included, the event rate is

\begin{multline}
    \frac{dR}{dE_R} = N_\text{Xe}\frac{\rho_{\chi}}{m_{\chi}} 
    \int_{E_\text{min}}^{E_\text{max}}\int_0^{\infty}\int_{v_\text{min}}^{\infty} dv\, 
	dE_\text{Pb}\, dE_R\,\times \\
    v f(v) \left(n_\text{Pb} \frac{d\sigma_\text{Pb} }{dE_R} L \right) \frac{d\sigma_\text{Xe}}{dE_R}(v_{f}) \Theta(v_f - v_\text{min}(E_R))\,.
\end{multline}
Here $v_f$ is the final velocity after colliding with lead, which implicitly depends on the lead recoil energy $E_\text{Pb}$, and $v_\text{min}(E_R)$ is the minimum velocity to produce a given energy xenon recoil.

\subsection{Shielding by Earth}

Once the DM has upscattered, it is of course possible for it to downscatter in the Earth before reaching the detector. To account for this, we model the Earth as being composed purely of silicon, and compute the mean free path of a $\chi_2$ particle passing through silicon. We compute the probability that a DM particle will not downscatter between where it upscatters and the position of the detector, and factor this into the event rate. To first order, this amounts to restricting the length $L$, used in the previous sub-section, to be less than the mean free path for scattering on silicon. At large cross sections, this suppresses the upscattering probability relative to what it would be if we had used the aforementioned range of values for L. However, because the mean free path is inversely proportional to $\sigma_p$, and the probability of scattering on lead and xenon are both proportional to $\sigma_p$, the event rate still increases with cross section, just more slowly than it would if downscattering were not accounted for.

The restriction described in the previous paragraph ensures that the DM upscatters close enough to the detector to not subsequently downscatter on silicon. However, it is possible that the DM could have upscattered and downscattered multiple times before reaching the point of the final upscattering. This could slow the DM enough that it might not be able to upscatter again, or at least it could significantly change the recoil distribution. As shown below, this effect is only relevant for relatively large cross sections.

\section{Results}
\label{sec:results}

We consider two datasets to set our limits. The first is from a one-ton-year exposure of XENON1T, reported in Ref.~\cite{XENON:2018voc}. This analysis considered nuclear recoils with energies from 4.9 keV to 40.9 keV, and saw no significant excess over background. For this reference, we use the efficiency function reported therein.

The second is a XENON100 analysis which extends up to energies of 240 keV, reported in Ref.~\cite{XENON:2017fdd}. This work combines previously analyzed, low-energy data with a high-energy dataset above about 50 keV. The analysis in question is an effective-field-theory search, and the high energy considered stems from the fact that some of the operators considered produce strongly velocity- or momentum-dependent cross sections. This paper, in fact, set limits on inelastic DM, but only up to $\delta = 250$ keV. No significant excess was observed above background. We restrict ourselves to recoil energies above 50 keV, i.e. the high energy channel of this search, and assume a constant efficiency of 85\%, which is a decent fit to the reported acceptance and, if anything, very mildly conservative.

\begin{figure}[h!]
    \centering
    \includegraphics[width=\linewidth]{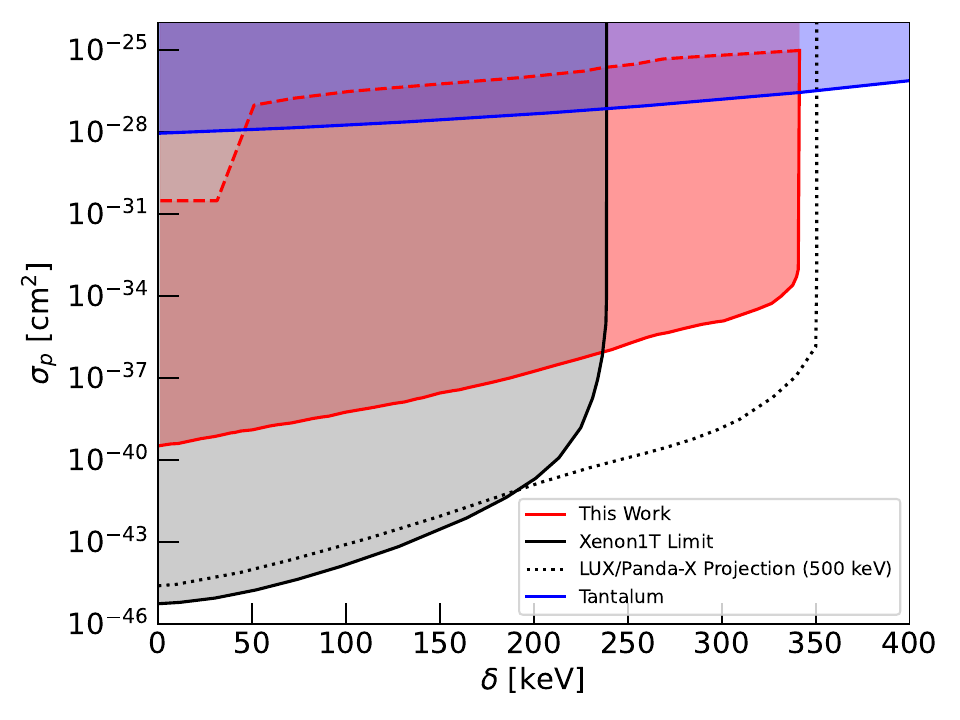}
	\caption{Present limit from XENON1T (shaded gray), taken from Ref.~\cite{Song:2021yar}, and a search using metastable tantalum~\cite{Lehnert:2019tuw}, compared to our results (red). Dotted black is a projection of what a xenon-based detector could probe by searching for nuclear recoils up to 500 keV~\cite{Bramante:2016rdh}. Dashed red is the maximum cross section that can be excluded by considering only DM that does not downscatter and re-upscatter on its way to the detector (see text for an explanation of the shape).}
    \label{fig:xenon1tlimits}
\end{figure}

Our results for XENON1T are shown in Fig.~\ref{fig:xenon1tlimits}, assuming $m_{\chi}$ = 1 TeV. Our limit extends up to a mass splitting of nearly 350 keV, whereas the current XENON1T limit extends only to about 250 keV. For comparison, we also show a projection from Ref.~\cite{Bramante:2016rdh}, which shows the sensitivity that a xenon-based detector could achieve by searching for recoils of up to 500 keV and seeing no significant background. Although our limit does not reach the lower cross section bounds in~\cite{Bramante:2016rdh}, we are
able to extend the exclusion region nearly as high in mass splitting without 
any need for higher energy data.

The dashed red curve in Fig.~\ref{fig:xenon1tlimits} is the maximum cross section that can be constrained by neglecting any DM that downscatters and subsequently upscatters again on its way to the detector. For large mass splittings, we consider only dark matter upscattering on lead. However, if the mass splitting is small enough, it could also upscatter on lighter nuclei such as silicon, which is much more abundant. This is the reason for the sharp bend in the dashed red curve around $\delta = 50$ keV: this marks the approximate mass splitting below which dark matter can upscatter on silicon in the Earth. Above roughly $\delta = 50$ keV, the dashed curve denotes the cross section where dark matter would upscatter multiple times on lead (potentially downscattering on silicon in between), while at lower mass splittings, it is the cross section where dark matter could upscatter multiple times on silicon. This dashed curve can be thought of as a very conservative estimate of the maximum cross section that can be constrained by the method presented here. Especially at small mass splittings, it could in fact be possible for DM to scatter many times and still produce a signal in the detector. Therefore, we extend the shaded red region above this dashed curve in lighter shading, and simply encourage caution when considering our results for such large cross sections.

The above limit extends higher in mass splitting than previous xenon-based analyses. However, there are other direct detection experiments, such as PICO-60~\cite{PICO:2015pux} and CRESST-II~\cite{CRESST:2015txj} that can probe mass splittings up to a bit over 400 keV. For this reason, we also consider another dataset, this one from XENON100~\cite{XENON:2017fdd}, which extends to much higher recoil energies, as mentioned above.

\begin{figure}[htb]
    \centering
    \includegraphics[width=\linewidth]{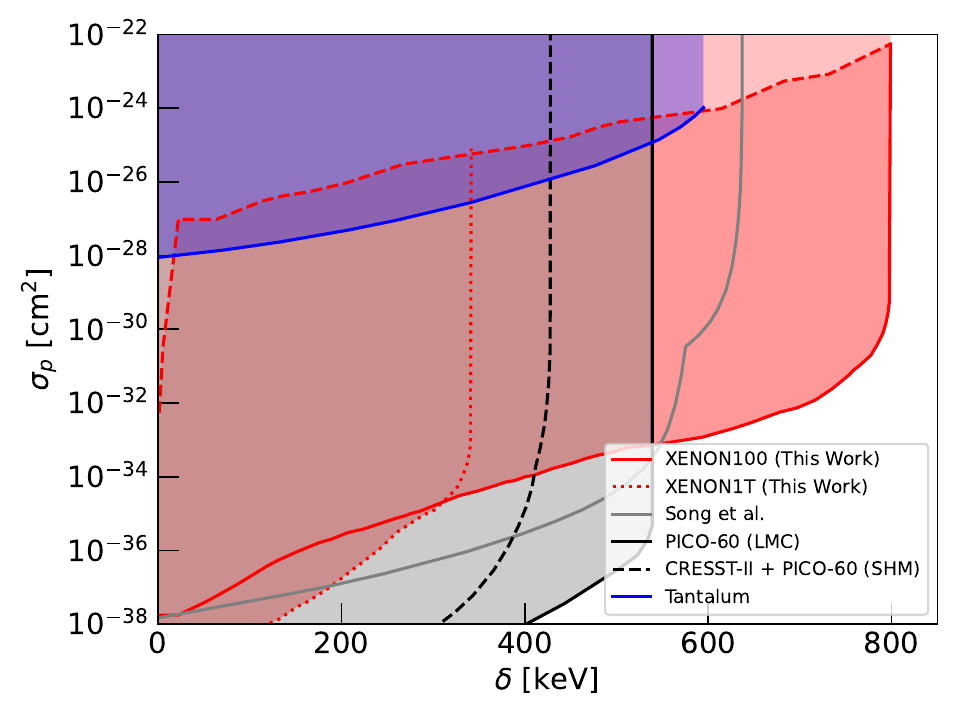}
    \caption{Our limit from XENON100 (solid red), compared to a compilation of direct detection limits assuming the Standard Halo Model (dashed black, from Ref.~\cite{Song:2021yar}), a limit from PICO-60~\cite{PICO:2023uff} assuming a contribution from the LMC (solid black, taken from Ref.~\cite{Graham:2024syw}), limits from Ref.~\cite{Song:2021yar} (gray), and a limit based on metastable tantalum~\cite{Lehnert:2019tuw}. Dashed red is as above. For reference, we also show our limit from XENON1T in dotted red.}
    \label{fig:xenon100limit}
\end{figure}

Our new limit derived from the XENON100 data is shown in Fig.~\ref{fig:xenon100limit}. The black shaded region is excluded by direct detection experiments; the region bounded by the dashed black curve assumes the Standard Halo Model and is reproduced from Ref.~\cite{Song:2021yar}, while the solid curve is taken from Ref.~\cite{Graham:2024syw} and represents a limit from PICO-60~\cite{PICO:2023uff} with the same assumption we use about the contribution of the LMC. The gray curve is a compilation of limits set in Ref.~\cite{Song:2021yar} based on measurements of radioactive decay in CaWO$_4$ and PbWO$_4$ crystals. And the shaded blue region is based on a search using metastable tantalum~\cite{Lehnert:2019tuw}. As before, the dashed red curve is the cross section above which DM may scatter many times on lead before reaching the detector. The result shown here extends to values of $\delta$ larger than those constrained by existing direct detection analyses by more than 200 keV. Due to the smaller exposure, the
bound is somewhat weaker than our XENON1T constraint at small $\delta$, but the extension to higher recoil energies allows us to constrain much larger mass splittings.

We note that there exists a constraint on much larger values of $\delta$ that arises from cosmic ray-DM scattering, but this result has only been presented for lower DM masses~\cite{Bell:2021xff}. Additionally, bounds from the gamma ray spectrum of metastable hafnium also constrain large mass splittings for cross sections above $10^{-21}$ cm$^2$~\cite{Alves:2023kek}. Finally, at cross sections above about $10^{-28}$ cm$^2$, Ref.~\cite{Acevedo:2025rqu} recently reported constraints on inelastic dark matter for mass splittings up to several MeV, based on scattering between dark matter and the gas cloud G2 near the galactic center. The strength of these limits varies depending on the assumed dark matter density profile, which assumes either the existence of a dark matter spike, or a generalized NFW profile with a relatively steep inner slope (compare to e.g. Refs.~\cite{2019JCAP...10..037D,Hussein:2025xwm}). Because several possible limits are reported depending on the choice of profile and target nucleus, we refer the reader to Ref.~\cite{Acevedo:2025rqu} for a detailed discussion.

\section{Conclusions}
\label{sec:conclusions}

In this work, we have studied the excitation of inelastic dark matter due to scattering in the Earth, followed by its deexcitation through scattering inside a detector. While past work has considered electromagnetic decays or electron interactions following such upscattering, we have performed a relatively model-independent analysis, relying only on nuclear scattering at both interaction points. The different kinematics between upscattering and downscattering allows us to set constraints, using existing data, on parameter space that cannot be constrained by considering only upscattering inside a detector. 

Although we have restricted our attention to xenon-based experiments, this formalism can be extended to essentially any direct detection experiment. It would be particularly interesting to consider detectors composed of much lighter nuclei: although such detectors are relatively insensitive to upscattering in the detector due to the lower center-of-mass kinetic energy, downscattering could still proceed unimpeded.

\acknowledgments
We are grateful to Jonas Frerick for helpful discussions. CVC was generously supported by Washington University in St. Louis through the Edwin Thompson Jaynes Postdoctoral Fellowship. FF would like to thank the Institut de F\'isica d'Altes Energies (IFAE) at the Universtitat Aut\`onoma de Barcelona (UAB) for its hospitality.

\bibliography{draft}

\end{document}